\documentstyle[10pt,epsf,epsfig,dp_delphititle,oldlfont]{dp_delphi}
\makeindex
\def\DpPaperGroup{EP}
\def\DpPaperRef{2000-127}
\def\DpDate{16 August 2000}
\def\DpAuthors{DELPHI Collaboration}
\def\DpSubmit{(Accepted by Physics Letters B)}
\def\DpTitle{{Search for R-parity violation with a \boldmath 
       $\bar{U} \bar{D} \bar{D}$ coupling at $\sqrt{s}=189$~GeV}}
\def\DpComment{ }
\def\DpEMail{ }

\begin{document}
\makeatletter
\newcount\@tempcntc
\def\@citex[#1]#2{\if@filesw\immediate\write\@auxout{\string\citation{#2}}\fi
  \@tempcnta\z@\@tempcntb\m@ne\def\@citea{}\@cite{\@for\@citeb:=#2\do
    {\@ifundefined
       {b@\@citeb}{\@citeo\@tempcntb\m@ne\@citea\def\@citea{,}{\bf ?}\@warning
       {Citation `\@citeb' on page \thepage \space undefined}}%
    {\setbox\z@\hbox{\global\@tempcntc0\csname b@\@citeb\endcsname\relax}%
     \ifnum\@tempcntc=\z@ \@citeo\@tempcntb\m@ne
       \@citea\def\@citea{,}\hbox{\csname b@\@citeb\endcsname}%
     \else
      \advance\@tempcntb\@ne
      \ifnum\@tempcntb=\@tempcntc
      \else\advance\@tempcntb\m@ne\@citeo
      \@tempcnta\@tempcntc\@tempcntb\@tempcntc\fi\fi}}\@citeo}{#1}}
\def\@citeo{\ifnum\@tempcnta>\@tempcntb\else\@citea\def\@citea{,}%
  \ifnum\@tempcnta=\@tempcntb\the\@tempcnta\else
   {\advance\@tempcnta\@ne\ifnum\@tempcnta=\@tempcntb \else \def\@citea{--}\fi
    \advance\@tempcnta\m@ne\the\@tempcnta\@citea\the\@tempcntb}\fi\fi}
 
\makeatother
\begin{titlepage}
\pagenumbering{roman}
\CERNpreprint{\DpPaperGroup}{\DpPaperRef} 
\date{{\small\DpDate}} 
\title{\DpTitle} 
\address{\DpAuthors} 
\begin{shortabs} 
\noindent
%
\noindent
Searches for pair production of gauginos and squarks in $e^{+} e^{-}$ collisions
at a centre-of-mass energy of 189 GeV have been performed on data corresponding
to an
integrated luminosity of 158 pb$^{-1}$ collected by the DELPHI detector at LEP.
The data were analyzed under the assumption of non-conservation of
\mbox{$R$-parity}
through a single dominant $\bar{U} \bar{D} \bar{D}$~coupling between squarks
and quarks.
Typical final states contain between 4 and 10 jets with or without additional
leptons.
No excess of data above Standard Model expectations was observed.
The results were used to constrain domains of the MSSM parameter space and
derive
limits on the masses of supersymmetric particles.

The following mass limits at 95\% CL were obtained from these searches:
\begin{itemize}
\item neutralino mass: $m_{\tilde{\chi^0_{1}}} \ge 32$ GeV
\item chargino  mass: $m_{\tilde{\chi^+_{1}}} \ge 94 $ GeV
\item stop and sbottom mass (indirect decay) with $\Delta M > 5$ GeV:

 $m_{\tilde{t_{1}}} \ge 74$ GeV, for $\Phi_{mix}=0$ rad

 $m_{\tilde{t_{1}}} \ge 59$ GeV,  for $\Phi_{mix}=0.98 $ rad

 $m_{\tilde{b_{1}}} \ge 72$ GeV, for $\Phi_{mix}=0$ rad.

\end{itemize}

The angle $\phi_{mix}$ is the mixing angle between left and right handed
quarks.
\end{shortabs}
\vfill
\begin{center}
\DpSubmit \ \\ 
\DpComment \ \\
\DpEMail \ \\
\end{center}
\vfill
\clearpage
\headsep 10.0pt
\addtolength{\textheight}{10mm}
\addtolength{\footskip}{-5mm}
\begingroup
%
\newcommand{\DpName}[2]{\hbox{#1$^{\ref{#2}}$},\hfill}
\newcommand{\DpNameTwo}[3]{\hbox{#1$^{\ref{#2},\ref{#3}}$},\hfill}
\newcommand{\DpNameThree}[4]{\hbox{#1$^{\ref{#2},\ref{#3},\ref{#4}}$},\hfill}
\newskip\Bigfill \Bigfill = 0pt plus 1000fill
\newcommand{\DpNameLast}[2]{\hbox{#1$^{\ref{#2}}$}\hspace{\Bigfill}}
%
\footnotesize
\noindent
\DpName{P.Abreu}{LIP}
\DpName{W.Adam}{VIENNA}
\DpName{T.Adye}{RAL}
\DpName{P.Adzic}{DEMOKRITOS}
\DpName{I.Ajinenko}{SERPUKHOV}
\DpName{Z.Albrecht}{KARLSRUHE}
\DpName{T.Alderweireld}{AIM}
\DpName{G.D.Alekseev}{JINR}
\DpName{R.Alemany}{VALENCIA}
\DpName{T.Allmendinger}{KARLSRUHE}
% ...
\normalsize
\endgroup
\titlefoot{Department of Physics and Astronomy, Iowa State
     University, Ames IA 50011-3160, USA
    \label{AMES}}
\titlefoot{Physics Department, Univ. Instelling Antwerpen,
     Universiteitsplein 1, B-2610 Antwerpen, Belgium \\
     \indent~~and IIHE, ULB-VUB,
     Pleinlaan 2, B-1050 Brussels, Belgium \\
     \indent~~and Facult\'e des Sciences,
     Univ. de l'Etat Mons, Av. Maistriau 19, B-7000 Mons, Belgium
    \label{AIM}}
\titlefoot{Physics Laboratory, University of Athens, Solonos Str.
     104, GR-10680 Athens, Greece
    \label{ATHENS}}
\titlefoot{Department of Physics, University of Bergen,
     All\'egaten 55, NO-5007 Bergen, Norway
    \label{BERGEN}}
\titlefoot{Dipartimento di Fisica, Universit\`a di Bologna and INFN,
     Via Irnerio 46, IT-40126 Bologna, Italy
    \label{BOLOGNA}}
\titlefoot{Centro Brasileiro de Pesquisas F\'{\i}sicas, rua Xavier Sigaud 150,
     BR-22290 Rio de Janeiro, Brazil \\
     \indent~~and Depto. de F\'{\i}sica, Pont. Univ. Cat\'olica,
     C.P. 38071 BR-22453 Rio de Janeiro, Brazil \\
     \indent~~and Inst. de F\'{\i}sica, Univ. Estadual do Rio de Janeiro,
     rua S\~{a}o Francisco Xavier 524, Rio de Janeiro, Brazil
    \label{BRASIL}}
\titlefoot{Comenius University, Faculty of Mathematics and Physics,
     Mlynska Dolina, SK-84215 Bratislava, Slovakia
    \label{BRATISLAVA}}
\titlefoot{Coll\`ege de France, Lab. de Physique Corpusculaire, IN2P3-CNRS,
     FR-75231 Paris Cedex 05, France
    \label{CDF}}
\titlefoot{CERN, CH-1211 Geneva 23, Switzerland
    \label{CERN}}
\titlefoot{Institut de Recherches Subatomiques, IN2P3 - CNRS/ULP - BP20,
     FR-67037 Strasbourg Cedex, France
    \label{CRN}}
\titlefoot{Now at DESY-Zeuthen, Platanenallee 6, D-15735 Zeuthen, Germany
    \label{DESY}}
\titlefoot{Institute of Nuclear Physics, N.C.S.R. Demokritos,
     P.O. Box 60228, GR-15310 Athens, Greece
    \label{DEMOKRITOS}}
\titlefoot{FZU, Inst. of Phys. of the C.A.S. High Energy Physics Division,
     Na Slovance 2, CZ-180 40, Praha 8, Czech Republic
    \label{FZU}}
\titlefoot{Dipartimento di Fisica, Universit\`a di Genova and INFN,
     Via Dodecaneso 33, IT-16146 Genova, Italy
    \label{GENOVA}}
\titlefoot{Institut des Sciences Nucl\'eaires, IN2P3-CNRS, Universit\'e
     de Grenoble 1, FR-38026 Grenoble Cedex, France
    \label{GRENOBLE}}
\titlefoot{Helsinki Institute of Physics, HIP,
     P.O. Box 9, FI-00014 Helsinki, Finland
    \label{HELSINKI}}
\titlefoot{Joint Institute for Nuclear Research, Dubna, Head Post
     Office, P.O. Box 79, RU-101 000 Moscow, Russian Federation
    \label{JINR}}
\titlefoot{Institut f\"ur Experimentelle Kernphysik,
     Universit\"at Karlsruhe, Postfach 6980, DE-76128 Karlsruhe,
     Germany
    \label{KARLSRUHE}}
\titlefoot{Institute of Nuclear Physics and University of Mining and Metalurgy,
     Ul. Kawiory 26a, PL-30055 Krakow, Poland
    \label{KRAKOW}}
\titlefoot{Universit\'e de Paris-Sud, Lab. de l'Acc\'el\'erateur
     Lin\'eaire, IN2P3-CNRS, B\^{a}t. 200, FR-91405 Orsay Cedex, France
    \label{LAL}}
\titlefoot{School of Physics and Chemistry, University of Lancaster,
     Lancaster LA1 4YB, UK
    \label{LANCASTER}}
\titlefoot{LIP, IST, FCUL - Av. Elias Garcia, 14-$1^{o}$,
     PT-1000 Lisboa Codex, Portugal
    \label{LIP}}
\titlefoot{Department of Physics, University of Liverpool, P.O.
     Box 147, Liverpool L69 3BX, UK
    \label{LIVERPOOL}}
\titlefoot{LPNHE, IN2P3-CNRS, Univ.~Paris VI et VII, Tour 33 (RdC),
     4 place Jussieu, FR-75252 Paris Cedex 05, France
    \label{LPNHE}}
\titlefoot{Department of Physics, University of Lund,
     S\"olvegatan 14, SE-223 63 Lund, Sweden
    \label{LUND}}
\titlefoot{Universit\'e Claude Bernard de Lyon, IPNL, IN2P3-CNRS,
     FR-69622 Villeurbanne Cedex, France
    \label{LYON}}
\titlefoot{Univ. d'Aix - Marseille II - CPP, IN2P3-CNRS,
     FR-13288 Marseille Cedex 09, France
    \label{MARSEILLE}}
\titlefoot{Dipartimento di Fisica, Universit\`a di Milano and INFN-MILANO,
     Via Celoria 16, IT-20133 Milan, Italy
    \label{MILANO}}
\titlefoot{Dipartimento di Fisica, Univ. di Milano-Bicocca and
     INFN-MILANO, Piazza delle Scienze 2, IT-20126 Milan, Italy
    \label{MILANO2}}
\titlefoot{Niels Bohr Institute, Blegdamsvej 17,
     DK-2100 Copenhagen {\O}, Denmark
    \label{NBI}}
\titlefoot{IPNP of MFF, Charles Univ., Areal MFF,
     V Holesovickach 2, CZ-180 00, Praha 8, Czech Republic
    \label{NC}}
\titlefoot{NIKHEF, Postbus 41882, NL-1009 DB
     Amsterdam, The Netherlands
    \label{NIKHEF}}
\titlefoot{National Technical University, Physics Department,
     Zografou Campus, GR-15773 Athens, Greece
    \label{NTU-ATHENS}}
\titlefoot{Physics Department, University of Oslo, Blindern,
     NO-1000 Oslo 3, Norway
    \label{OSLO}}
\titlefoot{Dpto. Fisica, Univ. Oviedo, Avda. Calvo Sotelo
     s/n, ES-33007 Oviedo, Spain
    \label{OVIEDO}}
\titlefoot{Department of Physics, University of Oxford,
     Keble Road, Oxford OX1 3RH, UK
    \label{OXFORD}}
\titlefoot{Dipartimento di Fisica, Universit\`a di Padova and
     INFN, Via Marzolo 8, IT-35131 Padua, Italy
    \label{PADOVA}}
\titlefoot{Rutherford Appleton Laboratory, Chilton, Didcot
     OX11 OQX, UK
    \label{RAL}}
\titlefoot{Dipartimento di Fisica, Universit\`a di Roma II and
     INFN, Tor Vergata, IT-00173 Rome, Italy
    \label{ROMA2}}
\titlefoot{Dipartimento di Fisica, Universit\`a di Roma III and
     INFN, Via della Vasca Navale 84, IT-00146 Rome, Italy
    \label{ROMA3}}
\titlefoot{DAPNIA/Service de Physique des Particules,
     CEA-Saclay, FR-91191 Gif-sur-Yvette Cedex, France
    \label{SACLAY}}
\titlefoot{Instituto de Fisica de Cantabria (CSIC-UC), Avda.
     los Castros s/n, ES-39006 Santander, Spain
    \label{SANTANDER}}
\titlefoot{Dipartimento di Fisica, Universit\`a degli Studi di Roma
     La Sapienza, Piazzale Aldo Moro 2, IT-00185 Rome, Italy
    \label{SAPIENZA}}
\titlefoot{Inst. for High Energy Physics, Serpukov
     P.O. Box 35, Protvino, (Moscow Region), Russian Federation
    \label{SERPUKHOV}}
\titlefoot{J. Stefan Institute, Jamova 39, SI-1000 Ljubljana, Slovenia
     and Laboratory for Astroparticle Physics,\\
     \indent~~Nova Gorica Polytechnic, Kostanjeviska 16a, SI-5000 Nova Gorica, Slovenia, \\
     \indent~~and Department of Physics, University of Ljubljana,
     SI-1000 Ljubljana, Slovenia
    \label{SLOVENIJA}}
\titlefoot{Fysikum, Stockholm University,
     Box 6730, SE-113 85 Stockholm, Sweden
    \label{STOCKHOLM}}
\titlefoot{Dipartimento di Fisica Sperimentale, Universit\`a di
     Torino and INFN, Via P. Giuria 1, IT-10125 Turin, Italy
    \label{TORINO}}
\titlefoot{Dipartimento di Fisica, Universit\`a di Trieste and
     INFN, Via A. Valerio 2, IT-34127 Trieste, Italy \\
     \indent~~and Istituto di Fisica, Universit\`a di Udine,
     IT-33100 Udine, Italy
    \label{TU}}
\titlefoot{Univ. Federal do Rio de Janeiro, C.P. 68528
     Cidade Univ., Ilha do Fund\~ao
     BR-21945-970 Rio de Janeiro, Brazil
    \label{UFRJ}}
\titlefoot{Department of Radiation Sciences, University of
     Uppsala, P.O. Box 535, SE-751 21 Uppsala, Sweden
    \label{UPPSALA}}
\titlefoot{IFIC, Valencia-CSIC, and D.F.A.M.N., U. de Valencia,
     Avda. Dr. Moliner 50, ES-46100 Burjassot (Valencia), Spain
    \label{VALENCIA}}
\titlefoot{Institut f\"ur Hochenergiephysik, \"Osterr. Akad.
     d. Wissensch., Nikolsdorfergasse 18, AT-1050 Vienna, Austria
    \label{VIENNA}}
\titlefoot{Inst. Nuclear Studies and University of Warsaw, Ul.
     Hoza 69, PL-00681 Warsaw, Poland
    \label{WARSZAWA}}
\titlefoot{Fachbereich Physik, University of Wuppertal, Postfach
     100 127, DE-42097 Wuppertal, Germany
    \label{WUPPERTAL}}
\addtolength{\textheight}{-10mm}
\addtolength{\footskip}{5mm}
\clearpage
\headsep 30.0pt
\end{titlepage}
%
\pagenumbering{arabic} 
\setcounter{footnote}{0} %
\large
%

\newcommand{\mm}      {\mbox{$ {\mathrm \mu}^+ {\mathrm \mu}^-             $}}
\newcommand{\tautau}  {\mbox{$ {\mathrm \tau}^+ {\mathrm \tau}^-           $}}
\newcommand{\LL}      {\mbox{$ {\mathrm l}^+ {\mathrm l}^-                 $}}
\newcommand{\qq}      {\mbox{$ {\mathrm q}\bar{\mathrm q}                  $}}
\newcommand{\vv}      {\mbox{$ {\mathrm \nu}\bar{\mathrm \nu}              $}}
\newcommand{\lv}      {\mbox{$ {\mathrm l}{\mathrm \nu}                    $}}
\newcommand{\ev}      {\mbox{$ {\mathrm e}{\mathrm \nu}                    $}}
\newcommand{\mv}      {\mbox{$ {\mathrm \mu}{\mathrm \nu}                  $}}
\newcommand{\tv}      {\mbox{$ {\mathrm \tau}{\mathrm \nu}                 $}}
\newcommand{\gam}     {\mbox{$ \gamma                                      $}}
\newcommand{\siminf}  {\mbox{$_{\sim}$ {\small {\hspace{-1.em}{$<$}}}       }}
\newcommand{\simsup}  {\mbox{$_{\sim}$ {\small {\hspace{-1.em}{$>$}}}       }}
\newcommand{\XO}      {$\widetilde{\chi}^0$}
\newcommand{\XOI}     {$\widetilde{\chi}_1^0$}
\newcommand{\XOIb}    {$\widetilde{\chi}_1^0$\ }
\newcommand{\XOII}    {$\widetilde{\chi}_2^0$}
\newcommand{\XOIIb}   {$\widetilde{\chi}_2^0$\ }
\newcommand{\XOIII}   {$\widetilde{\chi}_3^0$}
\newcommand{\XOIIIb}  {$\widetilde{\chi}_3^0$\ }
\newcommand{\XOIV}    {$\widetilde{\chi}_4^0$}
\newcommand{\XOIVb}   {$\widetilde{\chi}_4^0$\ }
\newcommand{\XOi}     {$\widetilde{\chi}_i^0$}
\newcommand{\XOj}     {$\widetilde{\chi}_j^0$}
\newcommand{\XOJ}     {$\widetilde{\chi}_i^0$}
\newcommand{\XOIJ}    {$\widetilde{\chi}_{1,2}^0$}

\newcommand{\XPI}{$\widetilde{\chi}_1^+$}
\newcommand{\XPIb}{$\widetilde{\chi}_1^+$\ }
\newcommand{\XMI}{$\widetilde{\chi}_1^-$}
\newcommand{\XMIb}{$\widetilde{\chi}_1^-$\ }
\newcommand{\XPM}{$\widetilde{\chi}^{\pm}$}
\newcommand{\XPk}{$\widetilde{\chi}_k^+$}
\newcommand{\XMl}{$\widetilde{\chi}_l^-$}

\newcommand{\SF}{$\widetilde{f}$}
\newcommand{\SQ}{$\widetilde{q}$}
\newcommand{\SNU}{$\widetilde{\nu}$}
\newcommand{\SLEP}{$\widetilde{l}$}
\newcommand{\SLEPP}{$\widetilde{l}^+$}
\newcommand{\SLEPM}{$\widetilde{l}^-$}
\newcommand{\STAU}{$\widetilde{\tau}$}
\newcommand{\STAUP}{$\widetilde{\tau}^+$}
\newcommand{\STAUM}{$\widetilde{\tau}^-$}
\newcommand{\SEL}{$\widetilde{e}$}
\newcommand{\SELP}{$\widetilde{e}^+$}
\newcommand{\SELM}{$\widetilde{e}^-$}
\newcommand{\SMU}{$\widetilde{\mu}$}
\newcommand{\SMUP}{$\widetilde{\mu}^+$}
\newcommand{\SMUM}{$\widetilde{\mu}^-$}

\newcommand{\epem}{$e^{+} e^{-}$}
\newcommand{\epemb}{$e^{+} e^{-}$\ }
\newcommand{\lum}{$\cal L$}

\newcommand{\mmub}{$\mu$\ }
\newcommand{\Mtb}{$M_2$\ }
\newcommand{\mo}{$m_0$}
\newcommand{\tanb}{$tan\beta$}

\newcommand{\Lijk}{$\lambda^{\prime \prime}_{ijk}$}
\newcommand{\Lpp}{$\lambda^{\prime \prime}$}
\newcommand{\Rpb}{$R_p$\hspace{-1.em}{\bf /}\ }
\newcommand{\Rp}{$R_p$\hspace{-1.em}{\rotatebox{-10}{/}}\ }
\newcommand{\UDD}{$\bar{U} \bar{D} \bar{D}$}

\newcommand{\Emiss}{\( \not \! {E} \) }
\newcommand{\Pt}{\mbox{$p_t$}}

\newcommand{\Ra}{$\rightarrow$\ }
\newcommand{\LRa}{$\Longrightarrow$\ }
\newcommand{\vs}{voie $s$}
\newcommand{\vt}{voie $t$}

\newcommand{\Zn}      {\mbox{$ {\mathrm Z}                               $}}
\newcommand{\Wp}      {\mbox{$ {\mathrm W}^+                               $}}
\newcommand{\Wm}      {\mbox{$ {\mathrm W}^-                               $}}
\newcommand{\Hp}      {\mbox{$ {\mathrm H}^+                               $}}
\newcommand{\WW}      {\mbox{$ {\mathrm W}^+{\mathrm W}^-                  $}}
\newcommand{\ZZ}      {\mbox{$ {\mathrm Z}{\mathrm Z}                 $}}
\newcommand{\GG}      {\mbox{$ {\mathrm \gamma}{\mathrm \gamma}            $}}
\newcommand{\HZ}      {\mbox{$ {\mathrm H}^0 {\mathrm Z}^0                 $}}
\newcommand{\GW}      {\mbox{$ \Gamma_{\mathrm W}                          $}}
\newcommand{\gs}      {\mbox{$ G_{\mathrm s}                          $}}
\newcommand{\gl}      {\mbox{$ G_{\mathrm L}                          $}}
\newcommand{\gr}      {\mbox{$ G_{\mathrm R}                          $}}
\newcommand{\gn}      {\mbox{$ G_{\mathrm \nu}                        $}}
\newcommand{\gmk}     {\mbox{$ G_{\mathrm k}                          $}}
\newcommand{\gml}     {\mbox{$ G_{\mathrm l}                          $}}

\newcommand{\nue}      {\mbox{$ \nu_ e                           $}}
\newcommand{\num}      {\mbox{$ \nu_\mu                          $}}
\newcommand{\nut}      {\mbox{$ \nu_\tau                         $}}

\newcommand{\Zg}      {\mbox{$ \Zn \gamma                                  $}}
\newcommand{\ecms}     {\mbox{$ \sqrt{s}                                    $}}
\newcommand{\ee}      {\mbox{$ {\mathrm e}^+ {\mathrm e}^-                 $}}
\newcommand{\eeWW}    {\mbox{$ \ee \rightarrow \WW                         $}}
\newcommand{\MeV}     {\mbox{$ {\mathrm{MeV}}                              $}}
\newcommand{\MeVc}    {\mbox{$ {\mathrm{MeV}}/c                            $}}
\newcommand{\MeVcc}   {\mbox{$ {\mathrm{MeV}}/c^2                          $}}
\newcommand{\GeV}     {\mbox{$ {\mathrm{GeV}}                              $}}
\newcommand{\GeVc}    {\mbox{$ {\mathrm{GeV}}/c                            $}}
\newcommand{\GeVcc}   {\mbox{$ {\mathrm{GeV}}/c^2                          $}}
\newcommand{\TeV}     {\mbox{$ {\mathrm{TeV}}                              $}}
\newcommand{\TeVc}    {\mbox{$ {\mathrm{TeV}}/c                            $}}
\newcommand{\TeVcc}   {\mbox{$ {\mathrm{TeV}}/c^2                          $}}
\newcommand{\MZ}      {\mbox{$ m_{{\mathrm Z}^0}                           $}}
\newcommand{\MW}      {\mbox{$ m_{\mathrm W}                               $}}
\newcommand{\GF}      {\mbox{$ {\mathrm G}_{\mathrm F}                     $}}
\newcommand{\MH}      {\mbox{$ m_{{\mathrm H}^0}                           $}}
\newcommand{\MT}      {\mbox{$ m_{\mathrm t}                               $}}
\newcommand{\GZ}      {\mbox{$ \Gamma_{{\mathrm Z}^0}                      $}}
\newcommand{\qqg}     {\mbox{$ {\mathrm q}\bar{\mathrm q}\gamma            $}}
\newcommand{\Wev}     {\mbox{$ {\mathrm{W e}} \nu_{\mathrm e}              $}}
\newcommand{\Zvv}     {\mbox{$ \Zn \nu \bar{\nu}                           $}}
\newcommand{\Zmm}     {\mbox{$ \Zn \mu^{+} \mu^-                           $}}
\newcommand{\Zee}     {\mbox{$ \Zn \ee                                     $}}
\newcommand{\ctw}     {\mbox{$ \cos\theta_{\mathrm W}                    $}}
\newcommand{\cwd}     {\mbox{$ \cos^2\theta_{\mathrm W}                  $}}
\newcommand{\cwq}     {\mbox{$ \cos^4\theta_{\mathrm W}                  $}}
\newcommand{\thw}     {\mbox{$ \theta_{\mathrm W}                          $}}
\newcommand{\gamgam}  {\mbox{$ \gamma \gamma                               $}}
\newcommand{\djoin}   {\mbox{$ d_{\mathrm{join}}                           $}}
\newcommand{\Erad}    {\mbox{$ E_{\mathrm{rad}}                            $}}
\newcommand{\mErad}   {\mbox{$ \left\langle E_{\mathrm{rad}} \right\rangle $}}

\newcommand{\sel}     {\mbox{$ \tilde{e}                                $}}
\newcommand{\sell}     {\mbox{$ \widetilde{e_L}                                $}}
\newcommand{\selr}     {\mbox{$ \widetilde{e_R}                                $}}
\newcommand{\smu}     {\mbox{$ \tilde{\mu}                              $}}
\newcommand{\smul}     {\mbox{$ \widetilde{\mu_L}                              $}}
\newcommand{\smur}     {\mbox{$ \widetilde{\mu_R}                              $}}
\newcommand{\stau}    {\mbox{$ \tilde{\tau}                             $}}
\newcommand{\staul}    {\mbox{$ \widetilde{\tau_L}                             $}}
\newcommand{\staur}    {\mbox{$ \widetilde{\tau_R}                             $}}
\newcommand{\snu}      {\mbox{$ \tilde{\nu}                                $}}
\newcommand{\snue}     {\mbox{$ \widetilde{\nu}_{e}                            $}}
\newcommand{\snum}     {\mbox{$ \widetilde{\nu}_{\mu}                          $}}
\newcommand{\snut}     {\mbox{$ \widetilde{\nu}_{\tau}                          $}}
\newcommand{\schi}    {\mbox{$ \tilde{\chi}                               $}}
\newcommand{\sph}    {\mbox{$ \tilde{p}                               $}}
\newcommand{\sm}       {\mbox{$ \tilde{m}                              $}}
\newcommand{\sfe}       {\mbox{$ \tilde{f}                              $}}
\newcommand{\mxo}       {\mbox{$ m_{\chi_o}                              $}}

\newcommand{\achi}    {\mbox{$ \tilde{\chi}                               $}}
\newcommand{\achiip}  {\mbox{$ \tilde{\chi^{+}_i}                         $}}
\newcommand{\achijp}  {\mbox{$ \tilde{\chi^{+}_j}                         $}}
\newcommand{\achijn}  {\mbox{$ \tilde{\chi^{-}_j}                         $}}
\newcommand{\achii}   {\mbox{$ \tilde{\chi^{0}_i}                         $}}
\newcommand{\achij}   {\mbox{$ \tilde{\chi^{0}_j}                         $}}
\newcommand{\achiap}    {\mbox{$ \tilde{\chi}^{+}_{1}                    $}}
\newcommand{\achibp}    {\mbox{$ \tilde{\chi}^{+}_{2}                       $}}
\newcommand{\achian}    {\mbox{$ \tilde{\chi}^{-}_{1}                       $}}
\newcommand{\achibn}    {\mbox{$ \tilde{\chi}^{-}_{2}                       $}}
\newcommand{\achia}    {\mbox{$ \tilde{\chi}^{0}_{1}                        $}}
\newcommand{\achib}    {\mbox{$ \tilde{\chi}^{0}_{2}                        $}}
\newcommand{\achic}    {\mbox{$ \tilde{\chi}^{0}_{3}                        $}}
\newcommand{\achid}    {\mbox{$ \tilde{\chi}^{0}_{4}                        $}}

\newcommand{\Ptau}    {\mbox{$ P_{\tau}                                    $}}
\newcommand{\mean}[1] {\mbox{$ \left\langle #1 \right\rangle               $}}
\newcommand{\mydeg}   {\mbox{$ ^\circ                                      $}}
\newcommand{\thetabar}{\mbox{$ \theta^*                                    $}}
\newcommand{\phibar}  {\mbox{$ \phi^*                                      $}}
\newcommand{\thetapl} {\mbox{$ \theta_+                                    $}}
\newcommand{\phipl}   {\mbox{$ \phi_+                                      $}}
\newcommand{\thetamin}{\mbox{$ \theta_-                                    $}}
\newcommand{\phimin}  {\mbox{$ \phi_-                                      $}}
\newcommand{\ds}      {\mbox{$ {\mathrm d} \sigma                          $}}
\newcommand{\emis}    {\mbox{$ E_{miss}                                  $}}
\newcommand{\ptr}     {\mbox{$ p_{\perp}                                   $}}
\newcommand{\ptrjet}  {\mbox{$ p_{\perp {\mathrm{jet}}}                    $}}
\newcommand{\Wvis}    {\mbox{$ {\mathrm W}_{\mathrm{vis}}                  $}}

\newcommand{\jjlv}    {\mbox{$ j j \ell                         $}}
\newcommand{\jj}      {\mbox{$ j j                              $}}
\newcommand{\lele}    {\mbox{$ \ell \bar{\ell}                  $}}
\newcommand{\eemis}   {\mbox{$ e \mu                            $}}
\newcommand{\emmis}   {\mbox{$ e e                               $}}
\newcommand{\mmmis}   {\mbox{$ \mu \mu                           $}}
\newcommand{\ttmis}   {\mbox{$ \tau \tau                         $}}
\newcommand{\ginv}    {\mbox{$ \gamma + "nothing"                        $}}
\newcommand{\jjll}    {\mbox{$ j j \ell \bar{\ell}                         $}}
\newcommand{\jjjj}    {\mbox{$ j j j j                                     $}}
\newcommand{\jjvv}    {\mbox{$ j j \nu \bar{\nu}                           $}}
\newcommand{\lvlv}    {\mbox{$ \ell \nu \ell \nu                           $}}
\newcommand{\llll}    {\mbox{$ \ell \ell \ell \ell                         $}}

\def\tht{\theta_{t}}
\def\st{\widetilde{t}}
\def\stl{\st_{1}}
\def\sth{\st_{2}}

\newcommand {\sutry} {supersymmetry}
\newcommand {\rp } {${R}_{p}$}
\newcommand {\rpv} {$\not \! {R}_{p}$}
\newcommand {\mtwo} {$M_{2}$}
\newcommand {\tb} {$\tan \beta$}
\newcommand {\mzero} {$m_{0}$}
\newcommand {\nump}{  \[ \sum_{n=0}^{n_{0}} \frac{(b+N)^{n}}{n!}\]  }
\newcommand {\denp}{  \[ \sum_{n=0}^{n_{0}} \frac{b^{n}}{n!} \] }

\newcommand{\lp}      {\mbox{$ \lambda'         $}}
\newcommand {\G} {$GeV/c^{2}$}
\newcommand {\M} {$MeV/c^{2}$}
\newcommand {\Gc} {$GeV/c$}
\newcommand {\Mc} {$MeV/c$}

\newcommand{\stp}     {\mbox{$ \tilde{t}                                   $}}
\newcommand{\sbt}     {\mbox{$ \tilde{b}                                   $}}
\def    \chipm         {\mbox{$\tilde\chi^{\pm}$}}   
\def    \chipma        {\mbox{$\tilde\chi^{\pm}_1$}}
\def    \chipa         {\mbox{$\tilde\chi^+_1$}}
\def    \chima         {\mbox{$\tilde\chi^-_1$}}
\def    \chipmb        {\mbox{$\tilde\chi^{\pm}_2$}}
\def    \chimpa        {\mbox{$\tilde\chi^{\mp}_1$}}
\def    \chimpb        {\mbox{$\tilde\chi^{\mp}_2$}}
\def    \chio          {\mbox{$\tilde\chi^0$}}
\def    \chioi         {\mbox{$\tilde\chi^0_i$}}
\def    \chioj         {\mbox{$\tilde\chi^0_j$}}
\def    \chioa         {\mbox{$\tilde\chi^0_1$}}
\def    \chiob         {\mbox{$\tilde\chi^0_2$}}
\def    \chioc         {\mbox{$\tilde\chi^0_3$}}
\def    \missEt      {\ifmmode{/\mkern-11mu E_t}\else{${/\mkern-11mu E_t}$}\fi}
\def    \missE          {\ifmmode{/\mkern-11mu E}\else{${/\mkern-11mu E}$}\fi}

\def\ap#1#2#3   {{\em Ann. Phys. (NY)} {\bf#1} (#2) #3.}
\def\apj#1#2#3  {{\em Astrophys. J.} {\bf#1} (#2) #3.}
\def\apjl#1#2#3 {{\em Astrophys. J. Lett.} {\bf#1} (#2) #3.}
\def\app#1#2#3  {{\em Acta. Phys. Pol.} {\bf#1} (#2) #3.}
\def\ar#1#2#3   {{\em Ann. Rev. Nucl. Part. Sci.} {\bf#1} (#2) #3.}
\def\cpc#1#2#3  {{Computer Phys. Comm.} {#1} (#2) #3.}
\def\err#1#2#3  {{\it Erratum} {\bf#1} (#2) #3.}
\def\epj#1#2#3  {{\em Eur. Phys. J.}{\bf#1} (#2) #3.}
\def\ib#1#2#3   {{\it ibid.} {\bf#1} (#2) #3.}
\def\jmp#1#2#3  {{\em J. Math. Phys.} {\bf#1} (#2) #3.}
\def\ijmp#1#2#3 {{\em Int. J. Mod. Phys.} {\bf#1} (#2) #3.}
\def\jetp#1#2#3 {{\em JETP Lett.} {\bf#1} (#2) #3.}
\def\jpg#1#2#3  {{\em J. Phys. G.} {\bf#1} (#2) #3.}
\def\mpl#1#2#3  {{\em Mod. Phys. Lett.} {\bf#1} (#2) #3.}
\def\nat#1#2#3  {{\em Nature (London)} {\bf#1} (#2) #3.}
\def\nc#1#2#3   {{\em Nuovo Cim.} {\bf#1} (#2) #3.}
\def\nim#1#2#3  {{Nucl. Instr. Meth.} {#1} (#2) #3.}
\def\np#1#2#3   {{Nucl. Phys.} {#1} (#2) #3.}
\def\npsup#1#2#3#4   {{\em Nucl. Phys.}{\bf #1} {\em (Proc. Suppl.)} 
{\bf#2} (#3) #4.}
\def\pcps#1#2#3 {{\em Proc. Cam. Phil. Soc.} {\bf#1} (#2) #3.}
\def\pl#1#2#3   {{Phys. Lett.} {#1} (#2) #3.}
\def\prep#1#2#3 {{Phys. Rep.} {#1} (#2) #3.}
\def\prev#1#2#3 {{Phys. Rev.} {#1} (#2) #3.}
\def\prl#1#2#3  {{\em Phys. Rev. Lett.} {\bf#1} (#2) #3.}
\def\prs#1#2#3  {{\em Proc. Roy. Soc.} {\bf#1} (#2) #3.}
\def\ptp#1#2#3  {{\em Prog. Th. Phys.} {\bf#1} (#2) #3.}
\def\ps#1#2#3   {{\em Physica Scripta} {\bf#1} (#2) #3.}
\def\rmp#1#2#3  {{\em Rev. Mod. Phys.} {\bf#1} (#2) #3.}
\def\rpp#1#2#3  {{\em Rep. Prog. Phys.} {\bf#1} (#2) #3.}
\def\sjnp#1#2#3 {{\em Sov. J. Nucl. Phys.} {\bf#1} (#2) #3.}
\def\spj#1#2#3  {{\em Sov. Phys. JEPT} {\bf#1} (#2) #3.}
\def\spu#1#2#3  {{\em Sov. Phys.-Usp.} {\bf#1} (#2) #3.}
\def\zp#1#2#3   {{\em Zeit. Phys.} {\bf#1} (#2) #3.}

\def\delnote#1#2#3#4 {{DELPHI} {#1}-{#2}~{\sc #3}~#4.}

\section{Introduction}
\subsection{The $R$-parity violating Lagrangian}
The most general way to write a superpotential, 
including the symmetries and particle 
content of the Minimal Supersymmetric extension of the Standard Model 
(MSSM)~\cite{review} is:
\begin{equation}
 W=W_{MSSM}+W_{RPV}
\end{equation}
where $W_{MSSM}$ represents interactions between MSSM particles 
consistent with $B-L$ conservation ($B =$~baryon number, $L =$~lepton number) 
and $W_{RPV}$ describes interactions violating $B$ or $L$ conservation~\cite{fayet}. 
This latter term of the superpotential can explicitly be written 
as\footnote{An additional fourth term in eq.\ref{rpv}, 
describing a bilinear coupling between the left handed lepton 
superfield and the up-type Higgs field, is assumed to be zero~\cite{dreiner97}.}~\cite{weinberg}:
\begin{equation}
{\lambda}_{ijk}L_iL_j{\bar E}_k +
{\lambda}^{\prime}_{ijk}L_iQ_j{\bar D}_k +
{\lambda}^{\prime \prime}_{ijk} {\bar U}_i{\bar D}_j{\bar D}_k
\label{rpv}
\end{equation}
where $i$, $j$ and $k$ are the generation indices; $L$ and ${\bar E}$ 
denote the left-handed doublet lepton and the right-handed singlet 
charge-conjugated lepton superfields respectively, whereas $Q$, ${\bar U}$ and ${\bar D}$ 
denote the left-handed doublet quark and the right-handed singlet charge-conjugated up- 
and down-type quark superfields;
${\lambda}_{ijk}$, ${\lambda}^{\prime}_{ijk}$
and ${\lambda}^{\prime \prime}_{ijk}$ are the Yukawa couplings. 
The first two terms violate $L$ conservation, and the third term $B$ 
conservation. 
Since ${\lambda}_{ijk} = -{\lambda}_{jik}$, 
${\lambda}^{\prime \prime}_{ijk} = -{\lambda}^{\prime \prime}_{ikj}$,
there are 9 ${\lambda}_{ijk}$, 27 ${\lambda}^{\prime}_{ijk}$ and
9 ${\lambda}^{\prime \prime}_{ijk}$ leading to 45 additional couplings.

One major phenomenological consequence of R-parity violation (\rpv) is that
the Lightest Supersymmetric Particle (LSP) is allowed to decay into 
standard fermions. This fact modifies the signatures of the supersymmetric 
particle production compared to the expected signatures in case of $R$-parity 
conservation. 
First, the LSP may be a charged sparticle, for example 
a chargino (this case is considered in this paper). 
Second, due to the LSP decay into fermions, multi-lepton and multi-jet 
topologies are expected. 
In this paper, searches for pair produced 
neutralinos~(\XOJ), charginos~(\XPM) and squarks~(\SQ) were performed 
under the hypothesis of $R$-parity violation with one single dominant
\UDD~coupling. 
The \UDD~terms couple squarks to quarks and the experimental signature of 
the \rpv~ 
events thus becomes multiple hadronic jets, in most of the cases 
without missing energy. These signatures with $R$-parity violation 
through \UDD~terms have been already performed by the other LEP2 experiments~\cite{alephl3}.

\subsection{Pair production of gauginos and squarks}

Pair production of supersymmetric particles in MSSM with \rpv ~is the same 
as \rp~ conserved pair production, since the \UDD~couplings are not present 
in the production vertex.

The mass spectrum and the pair production cross sections 
of neutralinos and charginos are fixed, in the analyses 
described in this paper, by the three parameters of the
 MSSM theory assuming GUT scale unification of gaugino masses:   
$M_2$, the SU(2) gaugino mass parameter at the electroweak scale, 
$\mu$, the mixing mass term of the Higgs doublets at the electroweak
scale and \tanb, the ratio of the vacuum expectation values of the two Higgs 
doublets. The cross section depends also on the common scalar mass at 
the GUT scale, m$_0$, due to selectron or sneutrino exchange in the t-channel 
for sufficiently low sfermions masses.    

Pair production of squarks ($\tilde{q}$) is also studied in this paper.
Here the cross-section mainly depends on the squark masses. 
In the case of the third generation, the left-right mixing angle enters in the 
production cross-section as well. 
In the squark analysis two cases are considered:
one with no mixing, the second with the mixing angle which gives the 
lowest production cross-section.

\subsection{Direct and indirect decays of gauginos and squarks}

The decay of the produced sparticles can either be direct or indirect.
In a {\sl direct decay} the sparticle decays directly or via a virtual sparticle exchange 
to standard particles through an \rpv\ vertex. In an {\sl indirect decay} 
the sparticle first decays through an \rp~ conserving vertex to a standard particle and an
on-shell sparticle, which then decays through an \rpv\ vertex. 
The squark analysis is done considering only the indirect decay channels
which are dominant for coupling values considered in the present studies.
  
\begin{figure}[htb]
\begin{center}
\epsfig{file=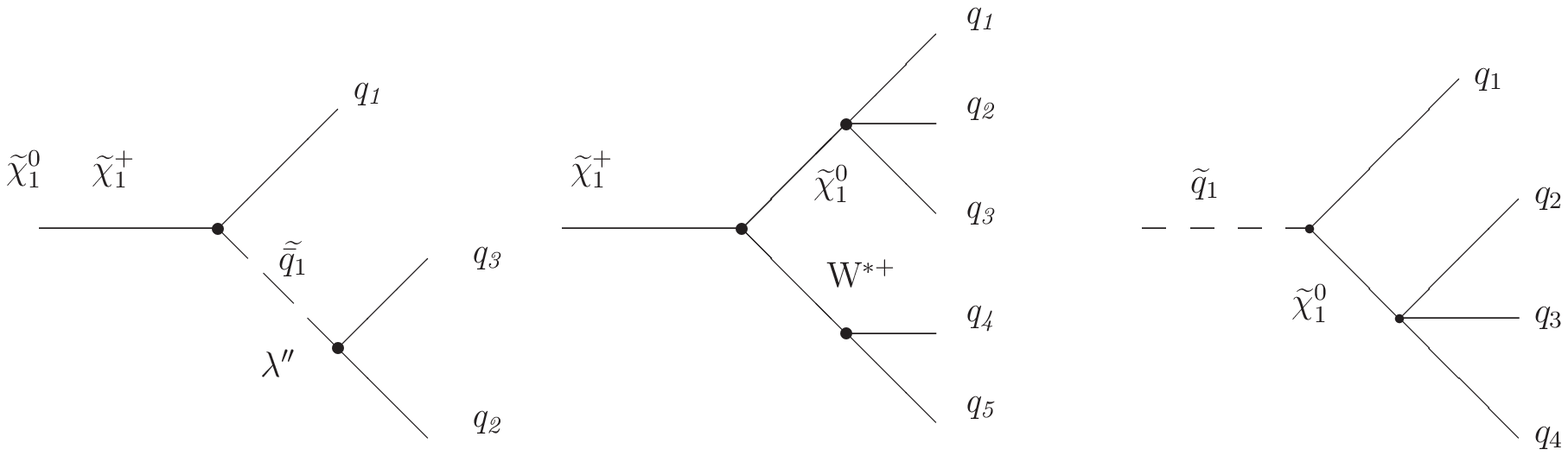,width=15cm}  
\vspace{-0.4cm}
\caption{\XOI, \XPI~direct decay (left), \XPI~(center) and \SQ~(right) indirect
  decay with a dominant \UDD~coupling. ${W^{*}}^+$ is an off-shell $W^+$ boson.}
\label{decay}
\end{center}
\end{figure}

Figure~\ref{decay} shows the direct and indirect 
decays of gauginos and the indirect decay of a squark via \UDD~couplings.


The most important features of these decays are the number of quarks    
in the final state which goes up to 10 for the indirect decay of two charginos.  
Table \ref{topos} displays the different event topologies from direct and 
indirect decays  through \UDD~couplings of different pair produced sparticles. 
The 6-, 8-, 10-jet topologies of table~\ref{topos} correspond to the decay 
diagrams in figure~\ref{decay}. 
\begin{table}[hbt]
\begin{center}
\begin{tabular}{ l l l } \hline
final states          & \multicolumn{1}{c}{direct}     & \multicolumn{1}{c}{indirect}  \\
                      & \multicolumn{1}{c}{decay of} &
                        \multicolumn{1}{c}{decay of} \\ 
\hline 
$4 j$                   & \SQ \SQ   &            \\
$6 j$                   & \XOI \XOI, \XOII \XOI, \XPI \XMI    &            \\
$8 j$                   &                          & \SQ \SQ \\
$10 j$                  &                          & \XPI \XMI \\
\hline
\end{tabular}
\caption{The multijet final states in neutralino, chargino and squark pair production when one \UDD~coupling is dominant. The leptonic decays of ${W^*}$ are not listed in these final states since 
only pure hadronic events are considered in this study.  
}
\label{topos}
\end{center}
\end{table}

\subsection{ \UDD~Couplings}

The \UDD~Yukawa coupling strength, corresponding to a squark decay into two 
quarks, can be bound from above by indirect limits.

Upper limits on \UDD~couplings come from Standard Model constraints with experimental measurements:

- double nucleon decays for $\lambda''_{112}$ couplings~\cite{sher}, 

- $n-\bar{n}$ oscillations for $\lambda''_{113}$~\cite{zwirner},

- $R_l =\Gamma_{had}(Z^0)/\Gamma_l (Z^0)$ for $\lambda''_{312},\lambda''_{313}, 
\lambda''_{323}$~\cite{srid2,bhatt}.

The upper limits on the other $\lambda''$ couplings do not come from experimental bounds.  
They are obtained from the requirement of perturbative unification 
at the GUT scale of $10^{16}$ GeV. This gives a limit of 1.25 for a sfermion mass 
of 100 GeV~\cite{sher,probir}.  
Upper limits on the \UDD~couplings are reported in table \ref{udd_coupl}.
\begin{table}[!htb]
\begin{center}
\begin{tabular}{ c c c c c c}
\hline
\ ijk\  &\   $\lambda_{ijk}'' \  $ & ijk &\   $\lambda_{ijk}'' \  $ & ijk &  
\ $\lambda_{ijk}'' \  $ \\
 $\lambda_{uds}''(112)$ & $ 10^{-6}$ & $\lambda_{cds}''(212)$ & $1.25$  
&  $\lambda_{tds}''(312)$ & $0.43$  \\
 $\lambda_{udb}''(113)$ & $ 10^{-5}$ & $\lambda_{cdb}''(213)$ & $1.25$  
&  $\lambda_{tdb}''(313)$ & $0.43$ \\
 $\lambda_{usb}''(123)$ & $1.25$ & $\lambda_{csb}''(223)$ & $1.25$ 
&  $\lambda_{tsb}''(323)$ & $0.43$  \\
\hline
\end{tabular}
\end{center}
\caption{Upper limits on the \UDD~Yukawa couplings 
in units of ($ m_{\tilde{f}}/100$ GeV), 
where $m_{\tilde{f}}$ is the appropriate squark mass~\cite{dreiner97}.
  }
\label{udd_coupl}
\end{table}    

Our analysis, which does not search for long lived sparticles in the detector (displaced vertices), has a limited sensitivity to weak coupling strengths.
The coupling strength dependence of the mean decay length of the LSP is given by ~\cite{dawson,dreiner-ross}:
\begin{equation}
L(\mathrm{cm}) = 0.1 \ (\beta \gamma) 
\left({m_{\tilde{f}}\over{100~\mathrm{GeV}}} \right)^4 
\left({1~\mathrm{GeV}\over{m_{\widetilde{\chi}}}}\right)^5 
{1\over{\lambda''^2}}
\label{gaulife}
\end{equation}
if the neutralino or the chargino is the  LSP with $\beta \gamma =  P_{\widetilde{\chi}}/m_{\widetilde{\chi}}$.
The typical lower limit of sensitivity for this analysis ($L$ \siminf 1~cm) is of the order of 
$10^{-4}$ ($10^{-3}$) in case of a \XO or a \XPM~ of 30 GeV (10 GeV), with a squark mass of 100 GeV.

For the generation of all the signals a $\lambda''_{212}$ coupling of the 
strength 0.1 was used. A different choice between $10^{-2}$ and 0.5 
would not change the neutralino decay topologies.   
The choice of this specific coupling was arbitrary, 
since all the analyses in this paper were coupling independent.
Searches for decays through specific $\lambda''$ couplings, 
leading to the production of one or several $b$ quarks, may indeed use 
the advantage of $b$-tagging techniques to reach higher sensitivities, but at 
the cost of lost generality. The aim of this paper was instead to perform a 
general coupling independent analysis for each of the search channels. 


\section{Data and MC samples}
The analysis was performed on the data corresponding to an integrated 
luminosity of 158~pb$^{-1}$ collected during 1998 by the DELPHI 
detector~\cite{delphidet} at centre-of-mass energies around 189 GeV.

The  contributions to the background coming from the 
Standard Model processes:  four-fermion final states (WW, ZZ) and 
\Zg\Ra $q \bar{q} (\gamma)$ were considered.
The contribution from $\gamma\gamma$ events after preselection was found to 
be negligible, due to the high detected energy fraction and multiplicities of 
the studied signals.
For the \Zg\Ra $ q \bar{q} \gamma$ backgrounds, the PYTHIA~\cite{pythia} generator was 
used whereas the four-fermion final states were generated with 
EXCALIBUR~\cite{excal}. 

To evaluate signal efficiencies, sparticle production was generated using 
SUSYGEN ~\cite{susygen}. All generated signal and background events were 
processed with the DELPHI detector simulation program (DELSIM). 

\section{Analyses}

\subsection{Topologies and analysis strategy}
The present study covers the search for \XOI, \XPI and \SQ \  pair production. 
The analysis of the different decay channels can be 
organized on the basis of the number of hadronic jets in the final state.

For each multijet analysis, the clustering of hadronic jets was performed 
by the {\it ckern} package\cite{ckern} based on the Cambridge clustering algorithm\cite{camjet}. 
The choice of this clustering algorithm was motivated by its good performance 
for configurations with a mixture of soft and hard jets, the expected case
for \UDD~events. Moreover, the algorithm provides a good resolution for the 
jet substructure which is present in \UDD~indirect decays. 
For each event, {\it ckern} provides all possible configurations between two 
and ten jets. The value of the variable $y_{i+1}$ (for $i$ between 1 and  9), 
that is the  transition value of the DURHAM resolution variable $y_{cut}$ for a given $i$, 
which changes the characterization of an event from an $i$ to an $i+1$ jet configuration, 
constitutes a powerful tool to identify the topologies in multijet 
signals. 

A neural network method was applied in order to distinguish signals from 
Standard Model background events. The SNNS \cite{snns} 
package was used for the training and validation of the neural networks. 
The training was done on samples of simulated background and 
signal. The exact configuration and input variables of 
each neural network depended on the search channel. Each neural 
network provided a discriminant variable which was used to 
select the final number of candidate events for each analysis. 

\subsection{Hadronic preselection}
Preselection of pure hadronic events was performed 
at the starting point of the gaugino  and squark analyses.

The following preselection criteria were applied for the gaugino (squark) analyses: 

\begin{itemize}
\item the charged multiplicity had to be greater or equal to 15 (20); 
\item the total energy from charged particles was required to be greater than 
      {\mbox{0.30 $\times$ \ecms}}, 
\item the total energy was required to be greater than
      {\mbox{0.55 (0.53) $\times$ \ecms}},
\item the total energy from neutral particles was required to be less than
      {\mbox{0.50 (0.47)$\times$ \ecms}}.
\end{itemize}

\begin{figure}[htb]
\begin{center}
\epsfig{file=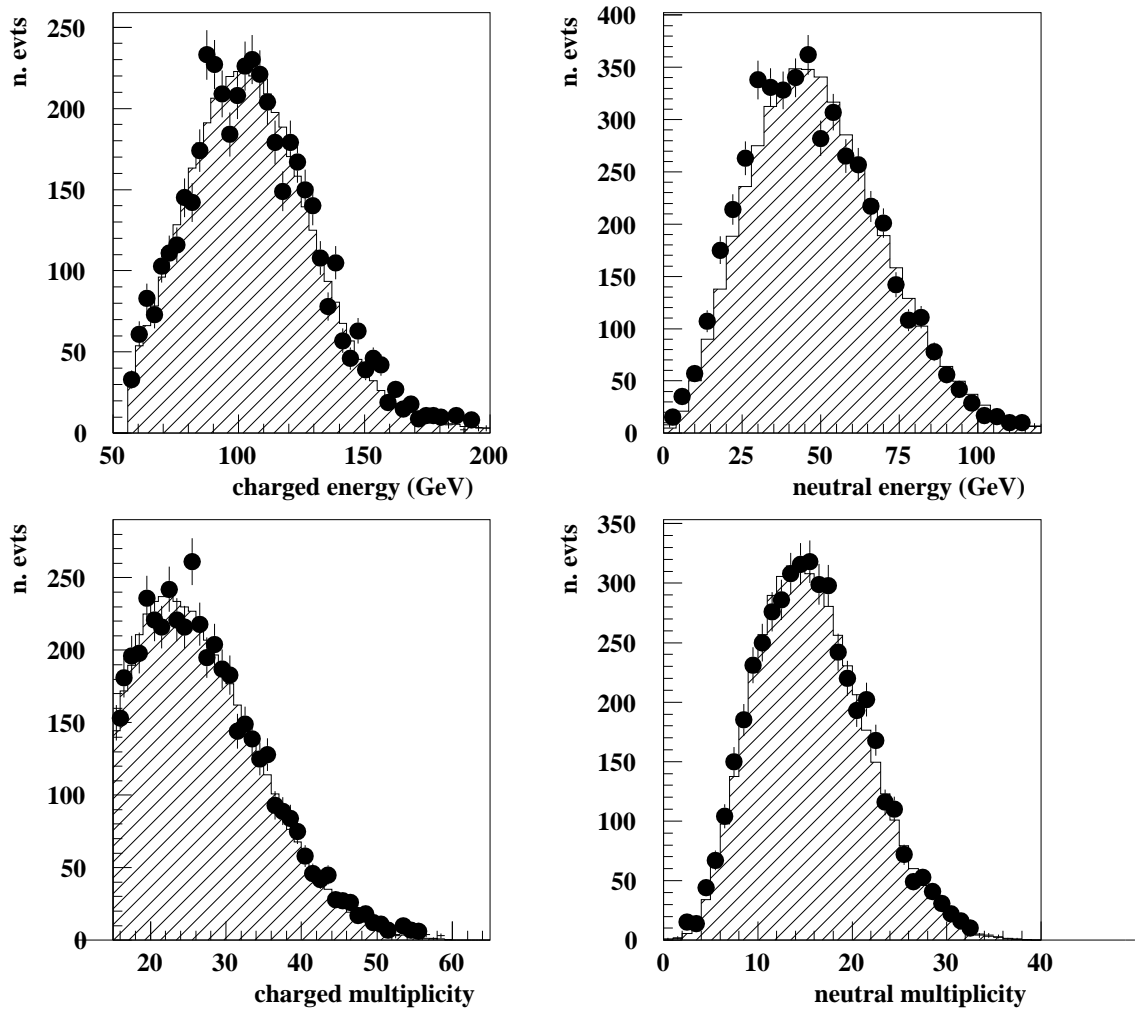,width=15cm}  
\vspace{-0.4cm}
\caption{Charged (upper left), neutral (upper right) energy distributions and
charged (lower left) and neutral (lower right) multiplicity distributions after hadronic preselection of gaugino analyses 
for data (black dots), expected SM background (hatched histograms).}
\label{agree}
\end{center}
\end{figure}

With these preselections most of the $\gamma \gamma$ background was 
suppressed. Tighter requirements  on charged multiplicity included in each 
analysis made this background negligible.
Therefore in what follows the main background  events will 
be the four-fermion events like \WW and the \Zg~ QCD events with hard gluon
radiation.
Signal efficiency at the level of hadronic preselection was between 80\% and 90\% 
for high and medium mass of pair-produced sparticles.
The preselection efficiency for the lowest neutralino mass was around 70\%.  
After the hadronic gaugino preselection the agreement between the  number of observed events (4722) in data 
and the number of expected events (4736) from  SM processes was rather good.
Figure~\ref{agree} shows the distributions of several variables after this hadronic preselection.    

\subsection{Charginos and neutralinos, 6- and 10-jet analyses}

To be efficient for all possible neutralino and 
chargino masses, the 6- (10-) jet analysis was divided 
into 3 (2) different mass windows.     

The signal selection in both channels was performed in 
two steps. 
First, we applied soft sequential criteria against mainly \Zg~ QCD events, except in the 
case of the low neutralino mass window:
\begin{itemize}
\item the effective centre-of-mass energy had to be greater than 
      {\mbox{150 GeV}},
\item the energy of the most energetic photon had to be less than 
      {\mbox{30 GeV}},
\item the sphericity had to be greater than 0.05, the thrust lower than 0.92 and -log(y$_3$) was 
required to be lower than 6.
\end{itemize}

Thereafter, a neural network method was used to select the signal against the \Zg~ QCD and
the four-fermion backgrounds. For each  analysis window a specific
neural network was trained.  
Topological variables used as inputs to the network were:
\begin{itemize} 
\item oblateness,
\item -log(y$_{n}$) with n=4 to 10,
\item minimum di-jet mass in 4-, 5- and 6-jet configurations,
\item energy of the least energetic jet $\times$ minimum di-jet angle in 4 and 5 jet configurations. 
\end{itemize}

The training was performed in a standard back-propagation manner
using the SNNS package \cite{snns}.
The network configuration had 13 input nodes, 13 hidden nodes and 3 output nodes.
The 3 output nodes correspond  to the  signal, the \Zg \  background and
the four-fermion background. 
This choice was motivated by the fact that we were looking for different 
signal topologies which were either similar to \Zg \ or to four-fermion events
depending on the analysis window.   

\subsubsection{Direct decay of \XOI \XOI\ or \XPI \XMI\ into 6 jets}

The 6-jet analysis was divided into 3 mass windows to 
take into account the magnitude of the gaugino boost depending on its mass: 
\begin{itemize}
\item window N1; low gaugino mass:    $ 10 \le m_{\tilde{\chi}} \le 30$ GeV, 
\item window N2; medium gaugino mass: $ 30 <  m_{\tilde{\chi}} \le 70$ GeV,
\item window N3; high gaugino mass:   $ 70 <  m_{\tilde{\chi}} \le 94$ GeV.
\end{itemize}

\begin{figure}[htb]
\begin{center}
\epsfig{file=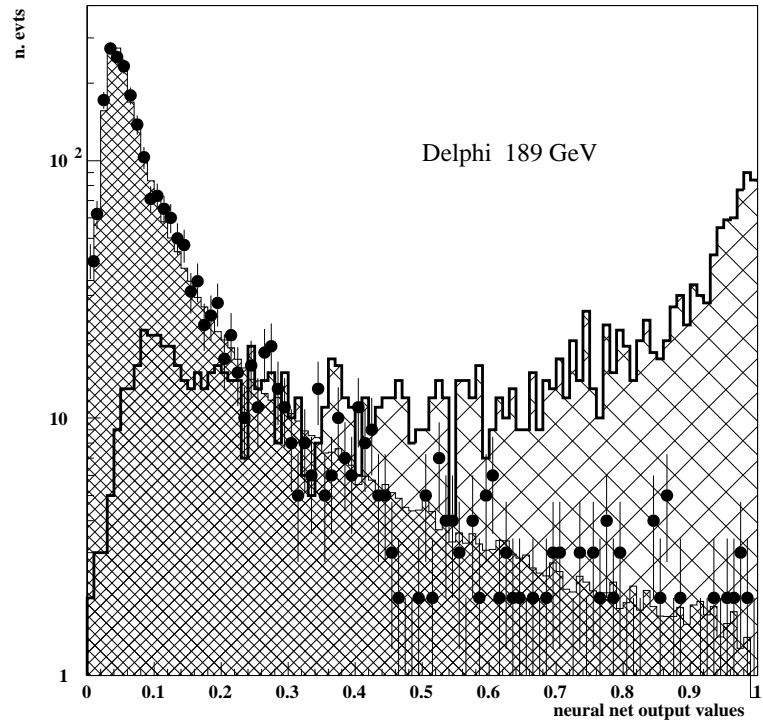,width=10cm}  
\vspace{-0.4cm}
\caption{Neural network signal output for data
(black dots), expected SM background (tight hatched) and the unweighted
signals (loose hatched) corresponding to the medium gaugino mass search N2.}
\label{nnw_n2}
\end{center}
\end{figure}

\begin{figure}[htb]
\begin{center}
\epsfig{file=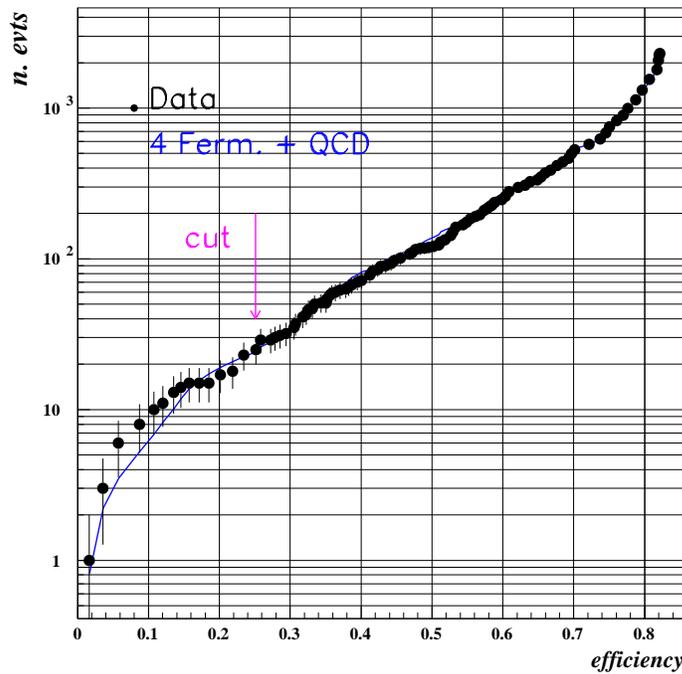,width=9cm}  
\vspace{-0.4cm}
\caption{Number of expected events (continuous line) data events (black dots)
versus signal efficiency for a 60 GeV neutralino mass
in the medium gaugino mass search N2. The arrow shows the efficiency 
corresponding to the working point.}
\label{effbkg_n2}
\end{center}
\end{figure}

The comparison between the number of expected SM background and the number of 
data events was performed for all neural network output values as is shown in Figure~\ref{nnw_n2}   
for  the medium gaugino N2 mass analysis window. 
Signal efficiencies were calculated  only from signal validation events 
(signal training events were not used at this level) for each neural network output value. 
Then the expected and obtained number of 
data events as a function of the signal efficiency was plotted as for example 
in Figure~\ref{effbkg_n2} for the  N2 analysis window.

No excess in the data appeared in these distributions, therefore a  
working point optimization on the neural network output was performed 
minimizing  the expected excluded cross-section as a function of the average 
signal efficiency of the mass window.
The working points of the neural network output were 
0.953, 0.852 and 0.966 for mass windows N1, N2 and N3 respectively. 
The corresponding 
signal efficiencies which increase with the neutralino mass 
were around  10-15\%, 25-30\% and 20-30\% for the 
mass windows N1, N2 and N3 respectively. To obtain signal efficiencies, the  
full detector simulation was performed on neutralino pair production 
with a 10 GeV step grid in the neutralino mass (10 to 94 GeV).
The statistical errors on the efficiencies was typically 2\%.
  
No excess of data over background was observed 
for any working point. The numbers of events seen and expected 
from backgrounds are shown in table~\ref{dmc_6}.

\begin{table}[hbt]
\begin{center}
\begin{tabular}{c c c c c } \hline
\ Window\   &\  Data \  &\  backgrounds\  &\  \Zg~background \ &\
four-fermion \ backgrounds \  \\
 N1    & 13 & 11.5 $\pm$ 0.4 & 10  &  1.5  \\
 N2    & 25 & 23.8 $\pm$ 0.5 & 2.6 & 21.2  \\
 N3    &  9 &  6.3 $\pm$ 0.3 & 0.4 &  5.9 \\
\hline
\end{tabular}
\caption{The numbers of events seen and expected from backgrounds for the three
mass windows of the 6-jet analysis.}
\label{dmc_6}
\end{center}
\end{table}

\subsubsection{Indirect decay of \XPI \  \XMI\  into 10 jets}

The 10-jet analysis was more sensitive to the mass difference 
between the chargino and the neutralino than to the neutralino mass. 
To take into account this mass difference we 
divided the 10-jet analysis into 2 windows:
\begin{itemize}
\item window C1; low chargino neutralino mass difference:  $\Delta M \le 10$ GeV, 
\item window C2; high chargino neutralino mass difference: $\Delta M > 10$ GeV.
\end{itemize} 

\begin{figure}[htb]
\begin{center}
\epsfig{file=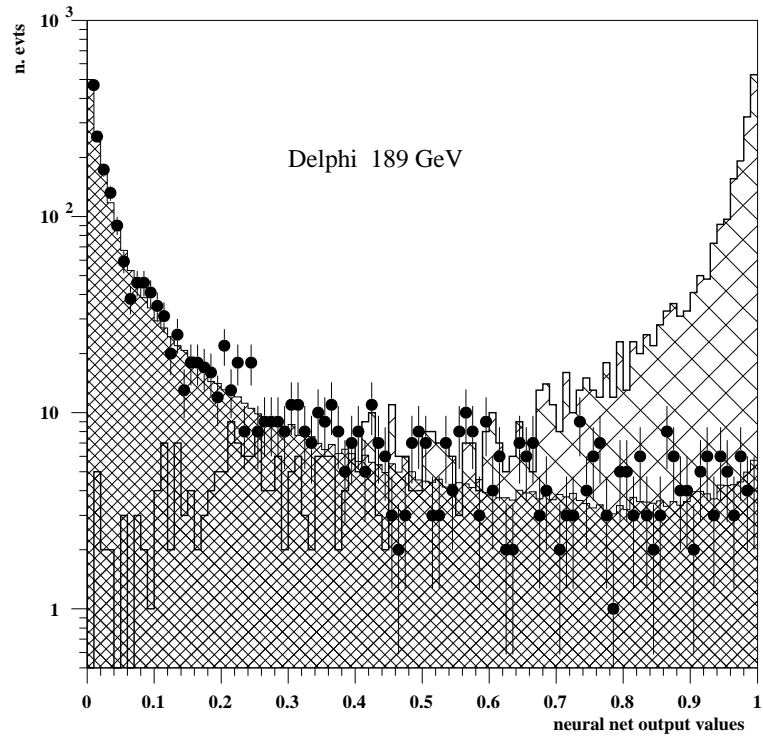,width=10cm}  
\vspace{-0.4cm}
\caption{Neural network signal output for data
(black dots), expected SM background (tight hatched) and the unweighted
signals (loose hatched) corresponding to the analysis applied in case of large $\Delta M$ between chargino and neutralino 
(window C2).}
\label{nnw_c2}
\end{center}
\end{figure}

\begin{figure}[htb]
\begin{center}
\epsfig{file=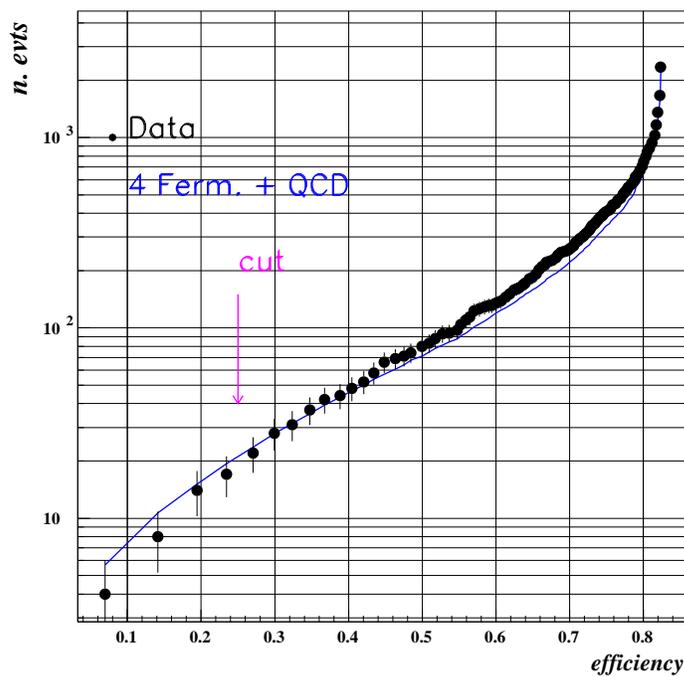,width=9cm}  
\vspace{-0.4cm}
\caption{Number of expected events (continuous line) data events (black dots) versus signal efficiency for a chargino mass of 80 GeV and a neutralino mass of
50 GeV in the analysis applied in case of large $\Delta M$ 
between chargino and neutralino 
(window C2). The arrow shows the efficiency 
corresponding to the working point.}
\label{effbkg_c2}
\end{center}
\end{figure}

The same neural network method was applied to select 10-jet events coming from indirect 
chargino decays. Two neural networks for the two different windows were produced.  
The distributions from expected SM events and data events were in good
agreement.  The neural network output of the C2 mass analysis
is given in Figure~\ref{nnw_c2} as an example.
Figure~\ref{effbkg_c2} shows the number of expected events 
and  data events as a function of the signal efficiency for the C2 mass window. 

The optimal working points 
have been found with the same procedure as for the 6-jet analysis.
The neural network output values were 
0.894 and 0.956 for two mass windows (C1 and C2).
The corresponding signal efficiencies were around  15-25\% and 10-50\% for the 
two mass windows.   
The statistical errors on the signal efficiency was 2\%.

In Figure~\ref{effbkg_c2} it can be seen that the background is not perfectly
reproduced by the simulation
in the high efficiency region dominated by $Z \gamma$ background, i.e. at the
preselection level. This region of high efficiency is not considered
in the final signal selection which is in the 10\%-50\% efficiency region.
The signal region is  mainly dominated by four-fermion background.
Therefore, an increase of
the uncertainty of the $Z \gamma$ background does not  drastically affect the
uncertainty on
the expected  background in the vicinity of the working point.


No excess was found in observed events compared 
to expected background for any  working point.
The numbers of events seen and expected 
from backgrounds are shown in table~\ref{dmc_10}.

\begin{table}[hbt]
\begin{center}
\begin{tabular}{c c c c c } \hline
\ Window\  &\ Data\  &\  backgrounds\  &\  \Zg~background\  &\  four-fermion
   \ backgrounds \  \\
 C1    & 28 & 25.3 $\pm$ 0.6 & 3.1  & 22.2 \\
 C2    & 18 & 21.0 $\pm$ 0.5 & 1.8  & 19.3 \\
\hline
\end{tabular}
\caption{The numbers of events seen and expected from backgrounds for the three
mass windows of the 10-jet analysis.}
\label{dmc_10}
\end{center}
\end{table}

\subsection{Squark 8-jet analysis}
Searches for squarks were performed in the case of indirect decays through a 
dominant $R$-parity violating \UDD~coupling. 
The final states in the indirect decay channel contain eight quarks of any 
flavour, but the topology of the signal strongly depends on the mass of 
the \XOI, through which the decay proceeds. SUSY signals were therefore 
simulated at different squark masses in the range 50-90 GeV with \XOI~masses 
between 10-80 GeV. The simulated decay actually used for the studies and efficiency evaluation 
was $\widetilde{b}$\Ra $b$ \XOI.

The general analysis methods    
based on a neural network background rejection were adopted for the analysis.
The analysis was aimed at a good sensitivity for $R$-parity violating \UDD~signals all over the plane of kinematically available squark and \XOI~masses.
First a general preselection, in addition to the one presented in section 3.2,
was made with the aim of a high general efficiency for the signal and at the 
same time a good rejection of low multiplicity hadronic background events.
The selection criteria were optimized for the 8-jet  squark analysis with
the following variables:
\begin{itemize}
\item the energy of the most energetic photon in the event had to be less than 45 GeV,
\item the missing momentum of the event had to be less than 76 GeV,
\item the oblateness of the event had to be less than 0.5.
\end{itemize}
A neural network was thereafter trained to calculate a discriminant variable 
for each event, in order to distinguish a possible signal from Standard Model 
background. The following quantities were used as input to the neural network:
\begin{itemize}
\item the total energy from neutral particles, the total event energy, the total number of charged particles, the energy of the most energetic photon in the event, the missing momentum of the event, the oblateness of the event,
\item  -log(y$_{n}$) with $n=2$ to $10$,
\item the reconstructed mass from a 5 constraint kinematic fit 
(the fifth constraint is the equal mass constraint on the di-jet masses) 
performed  on the 4 jet topology of the event and the $\chi^{2}$ value of this fit,
\item the minimum angle between two jets times the minimum jet energy from the 5 jet topology of the event.
\end{itemize}
Note that some of the input variables for the neural network were also used 
for the preselection, i.e. the preselection was used to 
eliminate the signal free regions and thereby unnecessary background from the analysis, whereas the neural network served to discriminate the signal from the background,
 in the remaining regions with overlapping values of the variables.
The final selection of candidate events was made based on the output value of 
the neural network. The 
working point optimization on the neural network output was performed 
minimizing  the expected excluded cross-section as a function of the average 
signal efficiency of the mass window.
No excess of data over Standard Model backgrounds was observed.
The numbers of events seen and expected 
from backgrounds are shown in table ~\ref{dmc_8}.

\begin{table}[hbt]
\begin{center}
\begin{tabular}{ c c c c} \hline
\  Data \   & \ backgrounds \  &\ \Zg~background\  &\ four-fermion
backgrounds \  \\     
    22 & 18.4 $\pm$ 0.7       & 3.8             & 14.6 \\
\hline
\end{tabular}
\caption{The numbers of events seen and expected from backgrounds for the three
mass windows of the 8-jet analysis.}
\label{dmc_8}
\end{center}
\end{table}

The signal efficiency was evaluated at each of the 30 evenly distributed 
simulated points in the plane of squark and neutralino masses and 
interpolated in the regions between. Efficiencies for the signal after the 
final selection range from 10-20\%, for small or large mass differences 
between squark and neutralino, up to 50\% for medium mass differences.
The statistical errors on signal efficiencies were typically 2\%.

\section{MSSM interpretation of the results} 
No excess was seen in the data with respect to the expected background
in any of the channels of these analyses.
Therefore, limits at 95\% confidence level on the cross-section of 
each process were obtained. Mass limits were derived
for supersymmetric  particles in the MSSM frame with \rpv. 
The cross-section ($\sigma_{95}$) that can be
excluded experimentally at 95\% confidence level, was calculated from 
data and SM event numbers obtained at the end of each analysis~\cite{pdg}.
\subsection{Chargino and neutralino multi-jet searches}

The excluded cross-sections, which is the $\sigma_{95}$ divided by the signal efficiency, 
are in the range [0.5,~0.7] pb, [0.2,~0.3] pb 
and [0.3,~0.4] pb for the N1, N2 and N3 neutralino analysis mass windows respectively 
and in the range [0.3,~0.6] pb and [0.1,~0.2] pb for the C1 and C2 indirect 
chargino decay analysis mass windows.   

The signal efficiency for any value of \XOI~and \XPM~masses
was interpolated using an efficiency grid determined with signal samples 
produced with the full DELPHI detector simulation. 
For typical values of tan$\beta$ and  $m_0$, a ($\mu$,{\mtwo})   point 
was excluded at 95\% confidence level if the signal cross-section times the 
efficiency at this point was greater than the cross-section ($\sigma_{95}$).

\begin{figure}[htb]
\begin{center}
\epsfig{file=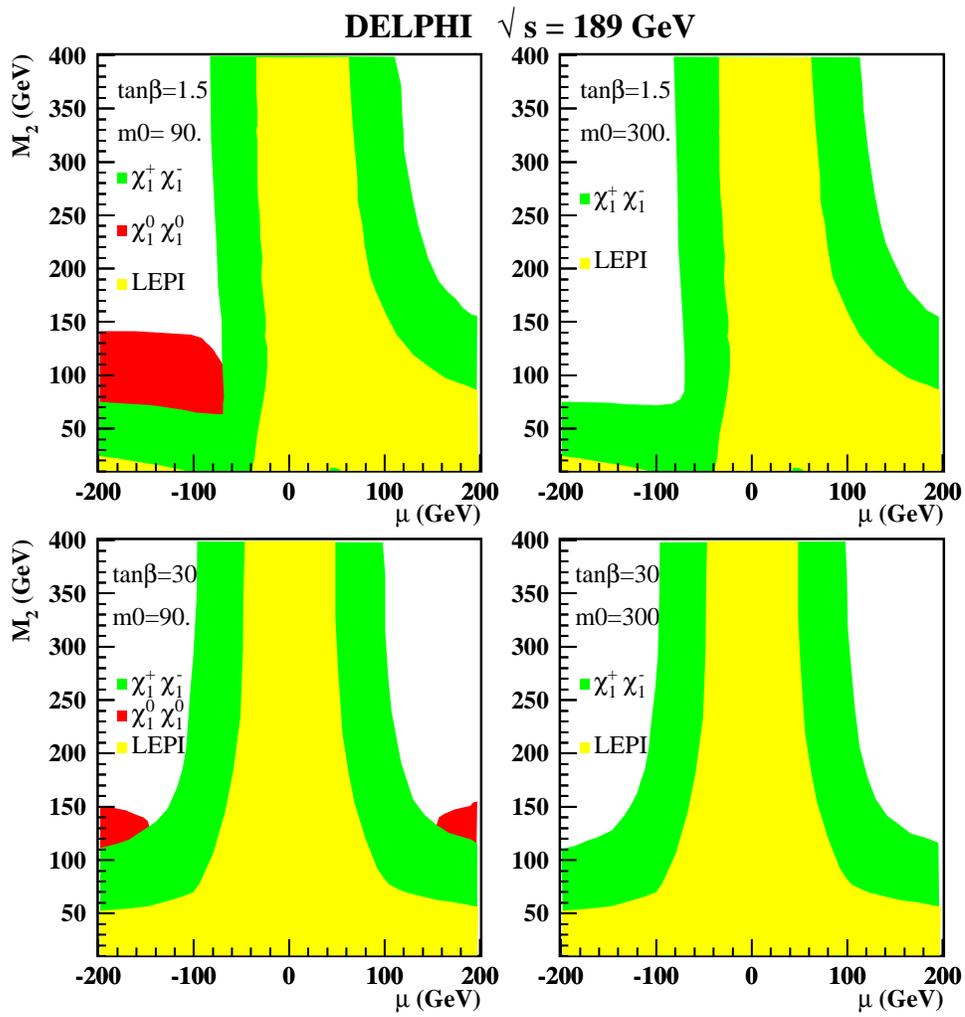,width=15cm}
\vspace{-1.0cm}
\caption{Exclusion plot in $\mu$, {\mtwo}  plane for $\chioa$ $\chioa$ and
  $\chipa$ $\chima$ production in the case of a dominant \UDD~$R$-parity violation coupling.
The 6- and 10-jets analyses are treated separately for this exclusion. The
shaded areas are excluded.}
\label{cha_lim}
\end{center}
\end{figure}

\begin{figure}[htb]
\begin{center}
\epsfig{file=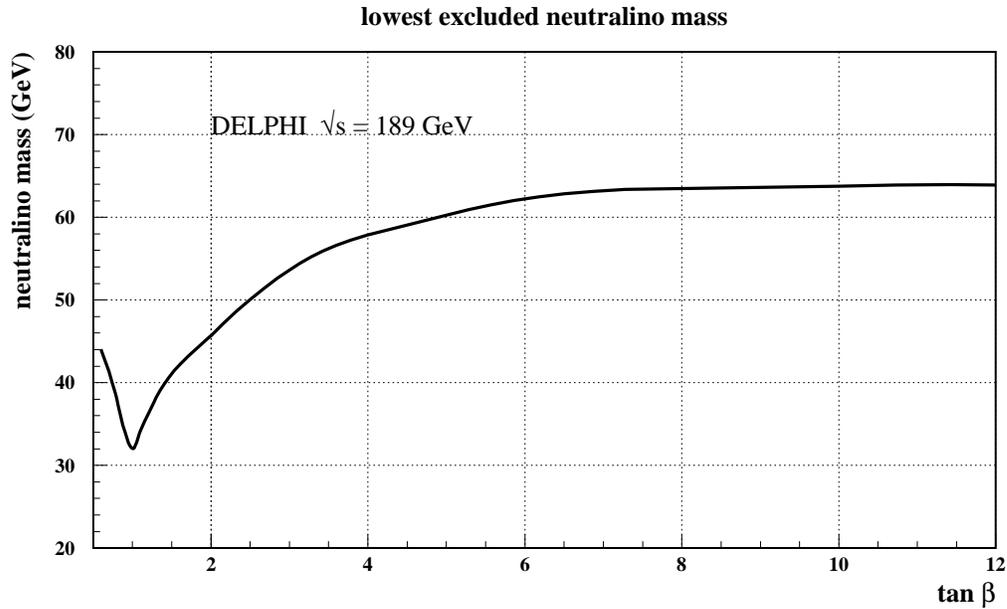,width=15cm}
\vspace{-0.5cm}
\caption{The excluded lightest neutralino mass as a function of tan $\beta$ 
at 95\% confidence level.}
\label{ne1_lim}
\end{center}
\end{figure}

Adding the 6-jet analysis (used for the direct decay of 
$\chipa \chima$ or $\chioa \chioa$) and the 10-jet 
analysis (used for indirect decay of $\chipa \chima$) results,
an exclusion contour in the $\mu$, {\mtwo} plane 
at 95\% confidence level was derived for different values of 
$m_0$ (90 and 300  GeV) and tan$\beta$ (1.5 and 30).
These exclusion contours in the $\mu$, {\mtwo} plane are shown in 
Figure~\ref{cha_lim}. In the exclusion plots the main contribution  comes 
from the study of the chargino indirect decays with the 10-jet analysis, 
due to the high cross-section. The 6-jet analysis becomes crucial 
in the exclusion plot for low tan$\beta$ value, low $m_0$ values and 
negative $\mu$ values. A 95 \% CL lower limits on the mass of lightest
neutralino and chargino  are obtained from  the $\mu$, {\mtwo} plane for 
different values of tan $\beta$ between 0.5 and 30 and for $m_0 = 500$ GeV.
The result on the lightest neutralino as a function of tan$\beta$ is shown 
in Figure~\ref{ne1_lim}. A lower limit on neutralino mass of 32 GeV is
obtained. The chargino is mainly excluded up to the 
kinematic limit at 94 GeV. 

\subsection{Indirect squark multi-jet searches}
\begin{figure}[htb]
\begin{center}
\epsfig{file=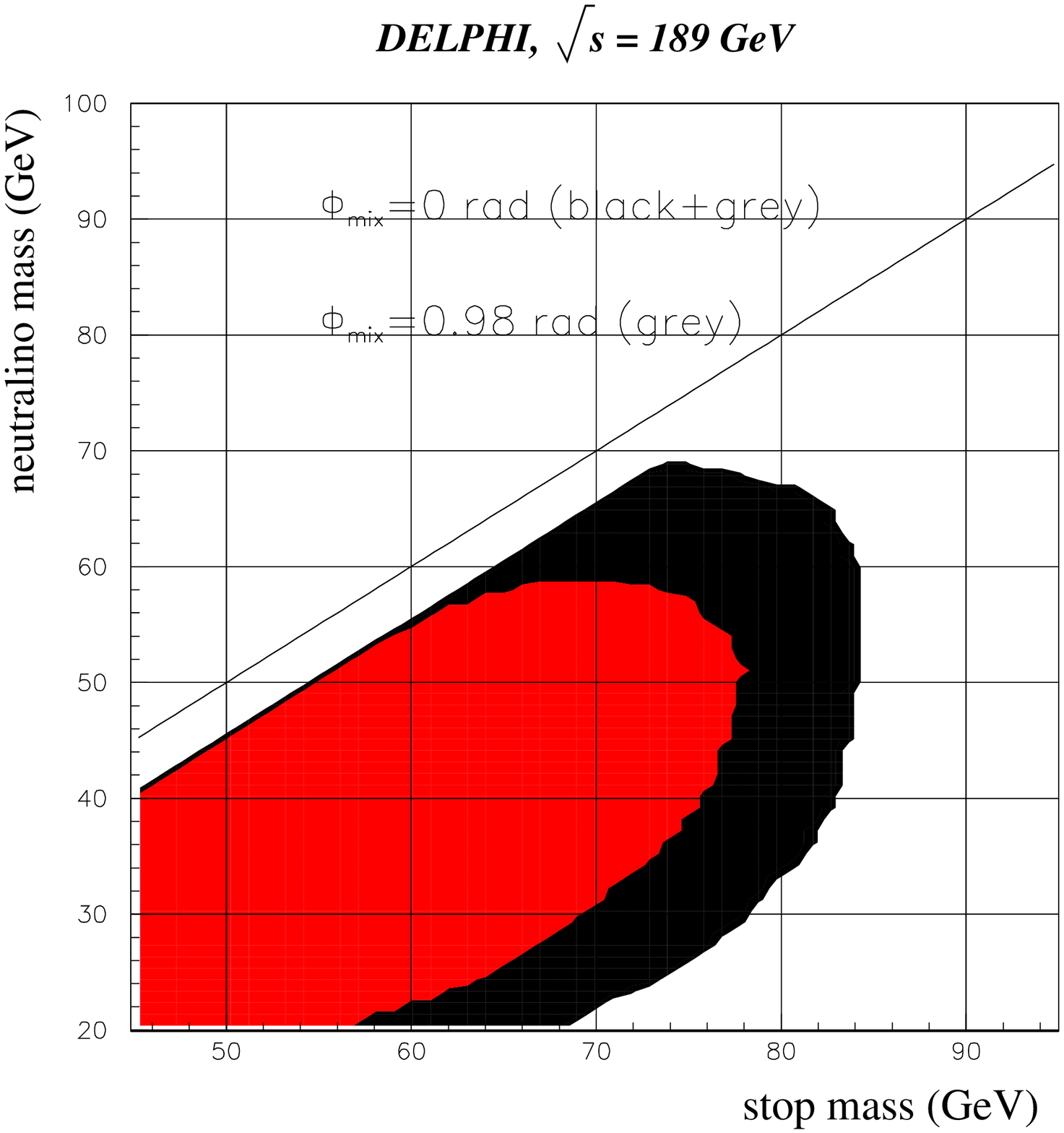,width=0.47\linewidth}
\epsfig{file=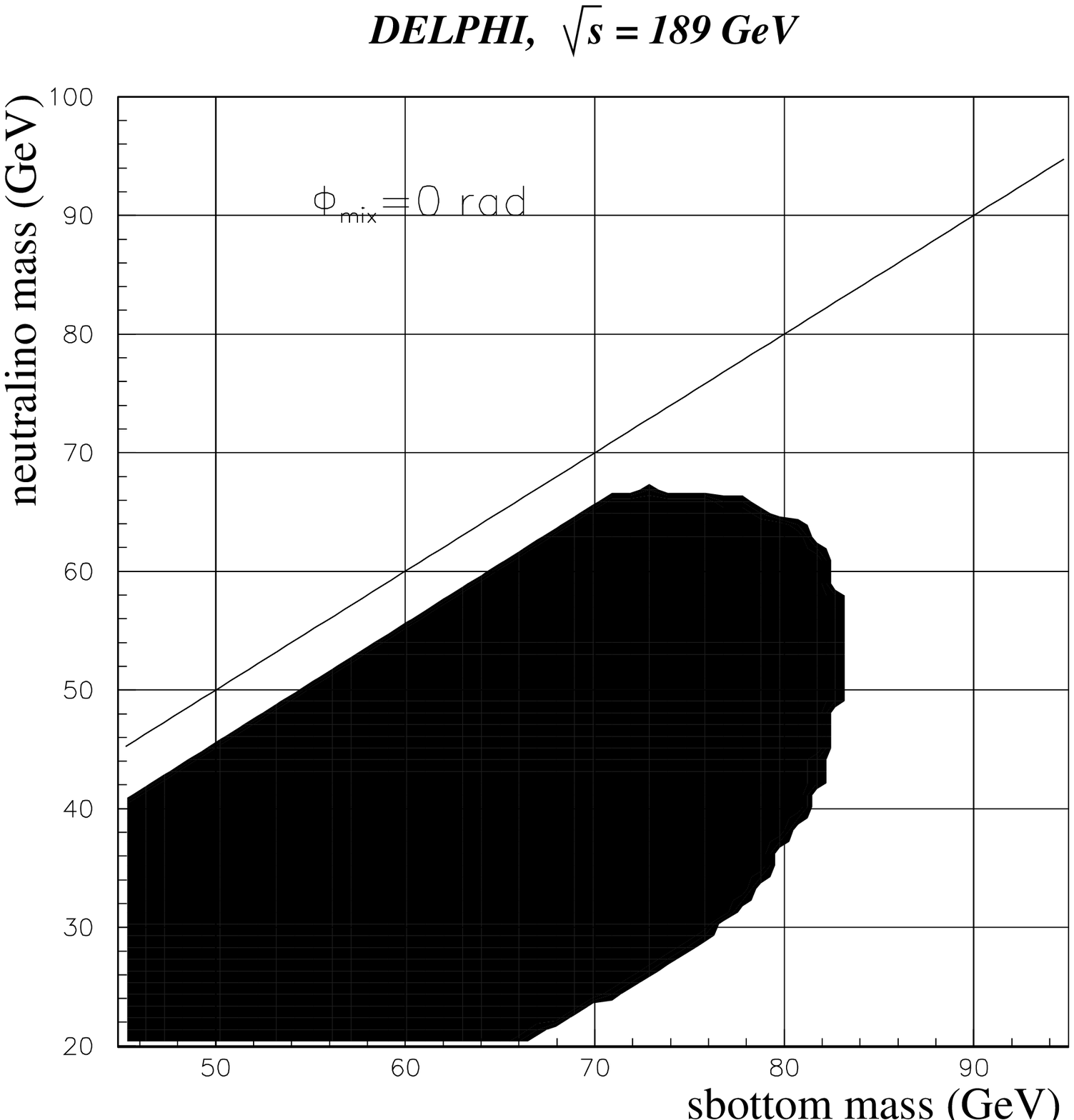,width=0.47\linewidth}
\vspace{-0.4cm}
\caption{Exclusion domains at 95\% confidence level in the 
$M(\tilde{\chi}^0_1)$, $M(\widetilde{q})$ plane for indirect squark decays in 
the case of a 100 \% branching ratio in the neutralino channel. The left plot 
shows the exclusion for a stop in the case of no mixing and with the mixing 
angle $\phi_{mix}$ which gives a minimum cross-section. For sbottom the minimum
cross-section is too low to extract any exclusion with the present analysis. 
The diagonal lines indicate the degenerate mass limit above which indirect 
squark decays are forbidden.}
\label{indexc}
\end{center}
\end{figure}

Exclusion domains were obtained by calculating $\sigma_{95}$ divided by the 
signal efficiency for each 1 GeV$\times$1 GeV bin in the neutralino mass
versus squark mass plane and comparing them to the cross-section for pair-produced 
squarks. 
The excluded cross-section varies between 0.2 and 0.9 pb depending on the efficiency. 
The resulting exclusion contours for stop and sbottom  can be 
seen in Figure~\ref{indexc}. A 100\% branching ratio of indirect decays in the 
neutralino channel was assumed for this exclusion. 
The mixing angle $0.98$ rad corresponds to the minimal lightest stop cross-section
due to a maximal decoupling from the Z boson.
  
By combining the exclusion contours from the squark 
searches with the constraint on the neutralino mass from the gaugino searches, 
lower bounds on the squark masses with 
$\Delta M > 5$ GeV are achieved. The lower mass limit on the stop is 74 GeV 
in the case of no mixing, and 59 GeV in the case of maximal Z-decoupling. The 
lower mass limit on the sbottom is 72 GeV in the case of no mixing.
For sbottom the minimum
cross-section is too low to extract any exclusion with the present analysis.

\section{Summary}
Searches for pair-produced gauginos and squarks, in the case of a single 
dominant $R$-parity violating \UDD~coupling, were performed on data collected 
by the DELPHI detector at a centre-of-mass energy of 189 GeV.
The analysis of the hadronic multijet final-states was performed by 
means of a neural network method and the results were interpreted within the 
framework of the MSSM.
No excess of data over the expected Standard Model events was found in 
any of the investigated search channels.
The result of the analysis implies the following lower mass limits, at a 
95 \% confidence level, on supersymmetric particles:
\begin{itemize}
\item neutralino mass: $m_{\tilde{\chi^0_{1}}} \ge 32$ GeV
\item chargino  mass: $m_{\tilde{\chi^+_{1}}} \ge 94 $ GeV
\item stop and sbottom mass (indirect decay) with $\Delta M > 5$ GeV:

 $m_{\tilde{t_{1}}} \ge 74$ GeV, for $\Phi_{mix}=0$ rad

 $m_{\tilde{t_{1}}} \ge 59$ GeV,  for $\Phi_{mix}=0.98 $ rad

 $m_{\tilde{b_{1}}} \ge 72$ GeV, for $\Phi_{mix}=0$ rad.
\end{itemize}
These mass limits were obtained under the following assumptions :
\begin{itemize}
\item One \UDD~term is dominant.
\item The limit on the neutralino and chargino masses were obtained for any 
m$_0$, tan $\beta$ values and for $-200< \mu <200$ GeV and $0 <$ {\mtwo} $<
400$ GeV.
\item The strength of the $\lambda''$ coupling was assumed to be greater 
than $10^{-3}$, limited by a mean LSP decay length smaller than 1 cm.
Smaller coupling strengths lead to a region between dominant $R$-parity
violation and $R$-parity conservation, which is not covered by these 
analyses.
\item stop and sbottom mass limits are valid for $\Delta M > 5$ GeV. They
were obtained for $\mu$ = -200 GeV and tan $\beta$ = 1.5. A branching ratio
of 100 \% into quark-neutralino was assumed.
\end{itemize}

\newpage

\subsection*{Acknowledgements}
\vskip 3 mm
 We are greatly indebted to our technical 
collaborators, to the members of the CERN-SL Division for the excellent 
performance of the LEP collider, and to the funding agencies for their
support in building and operating the DELPHI detector.\\
We acknowledge in particular the support of \\
Austrian Federal Ministry of Science and Traffics, GZ 616.364/2-III/2a/98, \\
FNRS--FWO, Belgium,  \\
FINEP, CNPq, CAPES, FUJB and FAPERJ, Brazil, \\
Czech Ministry of Industry and Trade, GA CR 202/96/0450 and GA AVCR A1010521,\\
Danish Natural Research Council, \\
Commission of the European Communities (DG XII), \\
Direction des Sciences de la Mati$\grave{\mbox{\rm e}}$re, CEA, France, \\
Bundesministerium f$\ddot{\mbox{\rm u}}$r Bildung, Wissenschaft, Forschung 
und Technologie, Germany,\\
General Secretariat for Research and Technology, Greece, \\
National Science Foundation (NWO) and Foundation for Research on Matter (FOM),
The Netherlands, \\
Norwegian Research Council,  \\
State Committee for Scientific Research, Poland, 2P03B06015, 2P03B1116 and
SPUB/P03/178/98, \\
JNICT--Junta Nacional de Investiga\c{c}\~{a}o Cient\'{\i}fica 
e Tecnol$\acute{\mbox{\rm o}}$gica, Portugal, \\
Vedecka grantova agentura MS SR, Slovakia, Nr. 95/5195/134, \\
Ministry of Science and Technology of the Republic of Slovenia, \\
CICYT, Spain, AEN96--1661 and AEN96-1681,  \\
The Swedish Natural Science Research Council,      \\
Particle Physics and Astronomy Research Council, UK, \\
Department of Energy, USA, DE--FG02--94ER40817. \\
\\
The financial support of STINT, The Swedish Foundation for International Cooperation in 
Research and Higher Education, and NFR, The Swedish Natural Science Research Council, is highly appreciated.


\newpage

\newpage

\newpage

\newpage


\newpage


\newpage


\end{document}